# Piezoelectricity in Monolayer Hexagonal Boron Nitride

Pablo Ares[1,2], Tommaso Cea[3], Matthew Holwill[1,2], Yi Bo Wang[1,2], Rafael Roldán[4], Francisco Guinea[1,3]*, Daria V. Andreeva[5], Laura Fumagalli[1,2]*, Konstantin S. Novoselov[1,6,7]*, and Colin R. Woods[1,2]

[1] Department of Physics & Astronomy, University of Manchester, Manchester, M13 9PL, UK
[2] National Graphene Institute, University of Manchester, Manchester, M13 9PL, UK
[3] Imdea Nanociencia, Faraday 9, Madrid, 28049, Spain
[4] Instituto de Ciencia de Materiales de Madrid, Sor Juana Inés de la Cruz 3, Madrid, 28049, Spain
[5] Department of Materials Science and Engineering, National University of Singapore, Singapore, 117575, Singapore
[6] Centre for Advanced 2D Materials, National University of Singapore, Singapore 117546, Singapore
[7] Chongqing 2D Materials Institute, Liangjiang New Area, Chongqing 400714, China

*Corresponding authors: francisco.guinea@manchester.ac.uk, laura.fumagalli@manchester.ac.uk konstantin.novoselov@manchester.ac.uk*

Two-dimensional (2D) hexagonal boron nitride (hBN) is a wide-bandgap van der Waals crystal with a unique combination of properties, including exceptional strength, large oxidation resistance at high temperatures and optical functionalities. Furthermore, in recent years hBN crystals have become the material of choice for encapsulating other 2D crystals in a variety of technological applications, from optoelectronic and tunnelling devices to composites. Monolayer hBN, which has no center of symmetry, has been predicted to exhibit piezoelectric properties, yet experimental evidence is lacking. Here, by using electrostatic force microscopy, we observed this effect as a strain-induced change in the local electric field around bubbles and creases, in agreement with theoretical calculations. No piezoelectricity was found in bilayer and bulk hBN, where the centre of symmetry is restored. These results add piezoelectricity to the known properties of monolayer hBN, which makes it a desirable candidate for novel electromechanical and stretchable optoelectronic devices, and pave a way to control the local electric field and carrier concentration in van der Waals heterostructures via strain. The experimental approach used here also shows a way to investigate the piezoelectric properties of other materials on the nanoscale by using electrostatic scanning probe techniques.

Piezoelectricity is an important property of non-centrosymmetric crystals that allows conversion of mechanical strain into electric field, and vice versa.[1] Recently, two-dimensional (2D) crystals have shown to be a unique platform to investigate and exploit such property for many reasons. First, they have the ability to sustain large strain (up to 10%) before rupture or plastic deformation,[2] while this is challenging to achieve in 3D crystals. Second, many crystals are found to be piezoelectric only when reduced to two-dimensionality. This is the case of semiconducting transition metal dichalcogenides, in which inversion symmetry is broken only in their 2D forms, as recently observed in single-layer $MoS_2$.[3] Furthermore, 2D crystals are likely to show areas of non-uniform strain near corrugations or bubbles that naturally form on substrates.[4] In such areas, strained-induced local charge densities, $\rho$, are expected to appear owing to the local variation in polarization, $P$, since $\rho(r) = -\nabla \cdot P(r)$.[5]

2D hexagonal boron nitride (hBN) is a van der Waals crystal with remarkable properties[2a, 6] and is an essential component of many new 2D technologies.[7] Recently, single-layer hexagonal boron nitride (hBN) has been theoretically predicted to be piezoelectric due to its broken inversion symmetry.[5, 8] Boron and Nitrogen atoms in hBN are arranged in a honeycomb lattice similarly as graphene, but the presence of different elements in the two sublattices of its unit cell makes it non-centrosymmetric. On the other hand, its bilayer and bulk counterparts recover the inversion symmetry and, therefore, no piezoelectricity is expected.[5, 8] Here we report experimental evidence of piezoelectricity in monolayer hBN by directly visualizing the strained-induced electric field in hBN single-layers using electrostatic force microscopy (EFM).[9] EFM images of monolayer hBN show enhanced electric contrast in correspondence of non-homogeneous strain areas around bubbles and creases. Such contrast vanishes on bilayer and few-layer hBN, as expected. We support our experimental findings with detailed theoretical simulations, solving the elasticity equations in a honeycomb lattice for deformations that mimic the observed bubbles and creases.

EFM is a non-contact scanning probe technique that maps the local electrostatic interaction between the tip and the sample under study.[9] We acquired EFM images by applying an ac voltage bias at frequency $\omega$ to the tip and measured the frequency shift of the cantilever at its mechanical resonance, detecting the electric response at the first and second harmonics, $\omega$ and $2\omega$, respectively.[10] We thus obtained two simultaneous EFM images: the electric image at $\omega$, which is proportional to the electric field on the surface,[10a, 10c, 11] and the dielectric image at $2\omega$, which depends only on the tip-sample capacitive interaction and the dielectric properties of the sample[10a, 10b] (see Experimental Section and Section 1 of Supporting Information (SI)). The latter was required here to investigate the impact of local capacitive variations on the electric image.

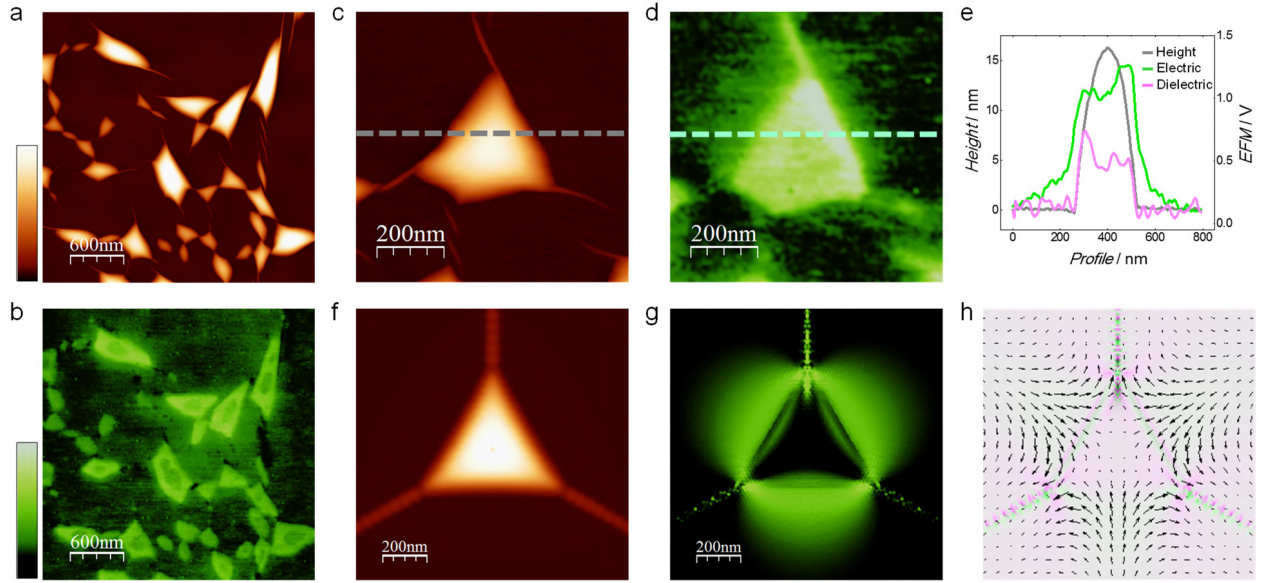

**Figure 1. Monolayer hBN on thick hBN crystals.** (a) Topography of a representative region with bubbles and creases. (b) Corresponding EFM electric image, showing enhanced contrast around and between bubbles. (c) Topography and (d) electric image of a triangular bubble. (e) Profiles along the lines in (c,d) and in the corresponding dielectric image (see Figure S4c in SI). A clear electric contrast extending hundreds of nm outside the bubble is observed (green line), while this is absent in the topography (grey line) and the dielectric image (pink line). (f) Simulated topography image of a triangular bubble in monolayer hBN and corresponding calculated (g) electric-field energy density and (h) polarization (vector field) and charge distribution (colormap). Color scales (from dark to bright): topography (both experimental and simulated) 20 nm; EFM 1.5 V; electric field energy density 1.5 µVÅ$^{-2}$; charge distribution -6×10$^{12}$ (green) to 6×10$^{12}$ (pink) e$^-$ cm$^{-2}$.

We applied EFM to monolayer, bilayer and few-layer hBN resting on top of either thick hBN crystals (thickness ~ 5-15 nm) or graphene on SiO$_2$/Si substrates, and not directly on top of SiO$_2$/Si substrates. This is because the use of a 2D crystal as bottom layer promotes the formation of non-uniform strain areas in the top layer. Such areas are observed around bubbles and creases filled with hydrocarbons that spontaneously appear in van der Waals heterostructures.[4] Furthermore, the use of thick hBN and graphene as bottom layers is beneficial to avoid the influence of localized charges trapped at the SiO$_2$ interface, facilitating the visualization of strained-induced electric fields in the monolayer hBN. We fabricated monolayer-hBN/thick-hBN and monolayer-hBN/graphene heterostructures using the standard dry transfer technique[12] (see Experimental Section).

Figure 1 focuses on the results obtained for monolayer hBN on top of thick hBN crystals. Figure 1a is a representative topography image of a monolayer hBN region with several bubbles and creases. The corresponding EFM electric image (Figure 1b) shows high contrast over the bubbles which is mainly due to the presence of molecules inside the bubbles. However, it also shows high contrast around the bubbles where the substrate is flat, extending hundreds of nanometers and connecting various bubbles. A different behavior is found in the corresponding dielectric image (see Figure S2c in SI). This shows the characteristic enhanced contrast over the bubbles caused by the material trapped inside,[10b] but no contrast around or between them. The absence of dielectric contrast outside the bubbles rules out that the origin of the electric contrast in these regions is a local change in dielectric properties. In particular, this allows discarding that the electric contrast outside the bubbles is caused by molecules that might be trapped below the monolayer. Electric and dielectric images obtained in other regions confirm these observations (see Figure S2d-f in SI). Furthermore, we did not observe such bright areas in flat regions of the monolayer in the absence of bubbles and creases (see Figure S3 in SI). This again implies that the electric field variations detected around them do not originate from molecules trapped at the hBN/hBN or hBN/SiO$_2$ interface. Close-up images confirmed these observations. Figure 1c shows the topography image of a triangular bubble. The regions in the immediate vicinity of the bubble are flat and featureless (see profile in Figure 1e), except for the creases from which the bubble originates. In contrast, the electric image (Figure 1d) clearly shows bright areas surrounding the bubble which are not found in the dielectric image (see Figure S4c in SI), as evidenced by the corresponding profiles (Figure 1e). We therefore conclude that such localized electric-field variations around the bubbles originate from the hBN being strongly strained in these areas. Additional images of triangular and elliptical bubbles further support this interpretation (see Figure S4d-f and S5 respectively in SI).

To support our experimental observations, we theoretically calculated the piezoelectric behavior of monolayer hBN in the presence of bubbles and creases that mimic those experimentally observed. Figure 1f-h plots the results of our simulations for the case of a triangular bubble similar to that experimentally observed in Figure 1c-d. The shape of the bubble is given by the equilibrium configuration of material trapped between a flat substrate and a 2D crystal attracted by van der Waals forces, as described in ref.[4]. For this shape we solved the discretized elasticity equations for the membrane with a honeycomb lattice (see Experimental Section and Sections 3 and 4 in SI), thus providing the strain tensor $u_{jk}$ and hence the piezoelectric induced polarization $P_i(r) = \sum_{jk}\gamma_{ijk}u_{jk}(r)$, being $\gamma_{ijk}$ the 3-rd rank piezoelectric tensor. For 2D crystals with D3h symmetry lying in the $xy$-plane (we choose $x$-direction parallel to the zigzag edge, and $y$-direction parallel to the armchair edge), the only non-zero independent coefficient is[5] $\gamma \equiv \gamma_{yyy} = -\gamma_{yxx} = -\gamma_{xyx} = -\gamma_{xxy}$. For the case of hBN and related 2D crystals with hexagonal symmetry, the polarization can be written as[5, 13] $P(r) = \gamma \mathcal{A}(r) \times \hat{z}$, where $\mathcal{A}(r) = (u_{xx}(r) - u_{yy}(r))\hat{x} - 2u_{xy}(r)\hat{y}$ has the form of the gauge field that appears in strained graphene.[14] We used the modern theory of polarization that exploits the geometrical properties of the Bloch wave-functions to obtain the electronic polarization,[15] a method that has been applied to non-centrosymmetric hexagonal nanotubes[16] and 2D crystals.[5, 13] In particular, it has been shown that it is possible to express the piezoelectric coefficient in terms of the valley Chern number.[5] For the case of interest here, the piezoelectric coefficient of hBN takes the simple form $\gamma = \eta \frac{e}{4\pi a_0} \mathcal{C}_{valley} \approx 2.91 \times 10^{-10}$ Cm$^{-1}$, where $\eta \approx 3.3$ is the electron-phonon coupling in hBN,[13] $a_0 = 1.44$ Å is the interatomic distance, $e$ is the elementary charge, and $\mathcal{C}_{valley} = \sum_\tau \tau \mathcal{C}_\tau = \text{sign}(\Delta)$ is the valley Chern number, where $\tau$ is the valley index, $\Delta \approx 5.97$ eV is the hBN bandgap, and $\mathcal{C}_\tau = \tau\, \text{sign}(\Delta)/2$. We refer to Experimental Section and Supporting Information for details in the numerical simulations steps. From these calculations we obtained the energy density generated by the piezoelectric effect (Figure 1g), and the spatial distribution of the electronic polarization $P(r)$ and the piezoelectric charge density $\rho(r)$ (Figure 1h).

Our simulations predict high contrast in energy density in the strained areas around the bubbles in correspondence of piezoelectric charge densities, in good agreement with our observations (Figure 1d). This can be understood as the EFM electric signal detects the electric field variations arising from local charge densities.[10a, 10c] Our simulations also predict a minimum in the center of the bubbles, which we experimentally detected in some of the EFM electric images (Figure 1b). However, the EFM contrast over the bubbles is affected by other important contributions in addition to piezoelectricity, including the dielectric properties of the molecules trapped inside, as already mentioned above, possible doping effects arising from them and topographic artefacts. Therefore, we limited our analysis to the experimental contrast observed outside the bubbles, where atomically flat and clean interfaces are present as a result of the self-cleansing mechanism of hBN crystals which pushes contamination away from the interfaces and gathers it into bubbles.[17] We theoretically analyzed the triangular bubble at different orientations with respect to the crystallographic axes (see Figure S15 in SI). We found that the energy density distribution does not depend on the bubble orientation, which is also consistent with our experimental observations. We note that a slight anisotropy in the contrast was experimentally detected around some bubbles. We attribute it to the asymmetries and imperfections of real bubbles as compared to the perfectly symmetric shapes of the bubbles that we simulated as well as to the asymmetric shape and scan angle of the AFM probe used in the experiments.

To better understand the experimental results, we investigated their dependence on the bottom layer used in our experiments. To this aim, we fabricated and measured heterostructures in which the monolayer hBN was transferred on a graphene layer instead of hBN crystals. Figure 2 plots representative experimental results for monolayer hBN on top of graphene. As shown in the topography image (Figure 2a), we found bubbles and creases with size and shape similar to those observed on hBN crystals. The corresponding EFM electric image (Figure 2b) also shows bright contrast in many flat areas extending hundreds of nanometers around and between bubbles and creases, as in the case of using hBN crystals as bottom layer. This contrast again vanishes in the corresponding dielectric image (see Figure S6c in SI), showing no features around or between bubbles. Images obtained in other regions confirm these observations (see Figure S6d-f in SI). Again, we did not observe any bright feature in the electric images in the absence of bubbles and creases (see Figure S7 in SI). We thus conclude that the local electric variations observed in our experiments do not originate in the bottom layer. They are the consequence of the strain in the hBN monolayer around bubbles and creases, in agreement with our theoretical analysis. As a further evidence, we found that for monolayer hBN on graphene, such bright areas tend to extend over larger regions and have higher directionality than those found on hBN crystals (see also SI). Figure 2d is a close-up electric image in one of these regions between two bubbles in monolayer hBN on graphene. The corresponding topography image (Figure 2c) reveals the presence of atomically thin creases (height ~ 3 Å) connecting the two bubbles. Figure 2e shows a high-resolution topographic image of a region around another bubble where atomically thin creases are clearly visible. These ultrathin creases, which are associated with strain concentration and release around the bubbles,[4, 18] produce additional strain and, therefore, generate an electric field that concentrates between bubbles. To support this conclusion, we simulated the case of two elliptical bubbles in the presence of atomically thin creases (Figure 2f-h), which mimic the ones observed in Figure 2c (see also Figure S17). The calculated electric field energy density image (Figure 2g) clearly exhibits higher contrast that extends between the two bubbles and matches the bright contrast observed in the EFM image in Figure 2d.

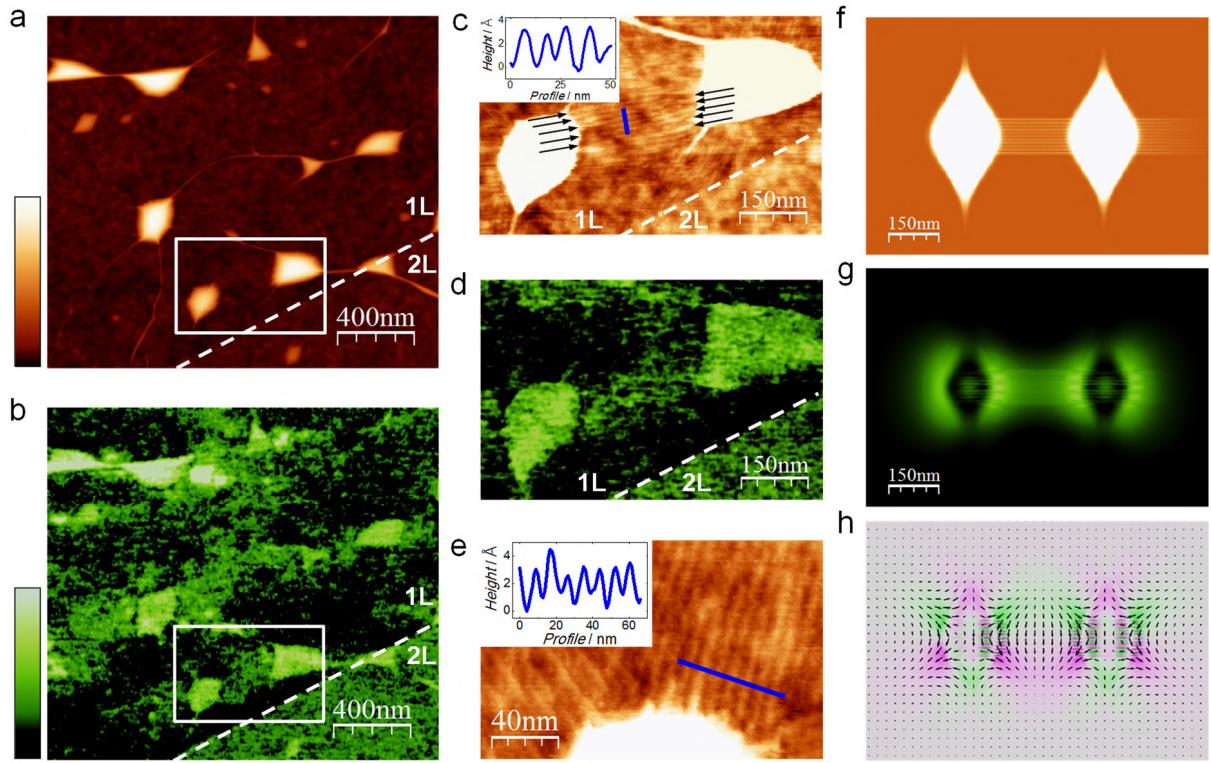

**Figure 2. Monolayer hBN on graphene**. (a) Topography of a representative region with bubbles and creases. The dashed line separates a monolayer area (main part of the image) from a bilayer terrace (bottom right corner). (b) Corresponding EFM electric image, showing enhanced contrast around and between bubbles in the monolayer. (c) Zoom in the region indicated by the rectangle in (a) showing atomically thin creases (marked by black arrows) between two bubbles. The inset is the topography profile taken along the blue line. (d) Corresponding electric image, showing enhanced contrast in correspondence of such ultrathin creases. (e) High-resolution topography image close to another bubble where ultrathin creases are clearly visible. The inset is the topography profile taken along the blue line. (f) Simulated topography of two elliptical bubbles connected by atomically thin creases and corresponding calculated (g) electric-field energy density, and (h) polarization (vector field) and charge distribution (colormap). Color scales (from dark to bright): topography 20 nm in (a), 2 nm in the rest; EFM 1.5 V; electric field energy density 75 µVÅ$^{-2}$; charge distribution -6×10$^{12}$ (green) to 6×10$^{12}$ (pink) e$^{-}$ cm$^{-2}$. The height of the bubbles in (c), (e), (f) was 20 nm, the color scale in these panels was adjusted to 2 nm to visualize the atomically thin creases.

To further support our experimental results, we fabricated and measured a series of control heterostructures for which no piezoelectricity is expected. In particular, we investigated the case of bilayer hBN, where the center of symmetry is restored, as well as the case of graphene, which is centrosymmetric because of the presence of the same element in the two sublattices of its unit cell. Figure 3 shows representative topography and EFM electric images of three heterostructures: bilayer hBN on thick hBN crystals, bilayer hBN on graphene, and graphene on graphene. Images were taken under the same experimental conditions as in Figure 2 and 3. In all the cases, we found bubbles of similar size and shape as those found in monolayer hBN (Figure 3a,c,e). The EFM electric images (Figure 3b,d,f) showed the usual high contrast over the bubbles due to the presence of trapped molecules, but no features outside the bubbles, in contrast with our observations in monolayer hBN (also see the dielectric images in Figure S8 in SI). We consistently found this behavior for all the bubbles in bilayer hBN irrespectively of the bottom layer as well as in graphene on graphene. Figure S9 and S10 in SI report additional images of different areas and bubbles in the control heterostructures.

Furthermore, Figure S11 in SI shows AFM and EFM images of few-layer hBN on both thick hBN crystals and graphene layers, in which we did not observe any sign of electric contrast outside the bubbles. All these experimental observations agree with the expected absence of piezoelectricity in bilayer and bulk hBN as well as in graphene. This strongly supports our interpretation of the electric contrast found in monolayer hBN as evidence of piezoelectricity. This also proves that other possible sources of contrast such as doping effects arising from the underlying layer, the presence of adsorbates or free carriers, which are not included in our simulations, are negligible, otherwise we would have detected them in these control measurements. Finally, we note that we found the presence of atomically thin creases around the bubbles in graphene over graphene (Figure S10d in SI) similarly as in monolayer hBN on graphene. However, no contrast was observed in the corresponding electric images. This clearly rules out that the strain-induced electric field detected in monolayer hBN in the presence of such ultrathin creases (Figure 2d) originates in the underlying graphene. It again supports our interpretation of such contrast as sign of piezoelectricity in the monolayer hBN.

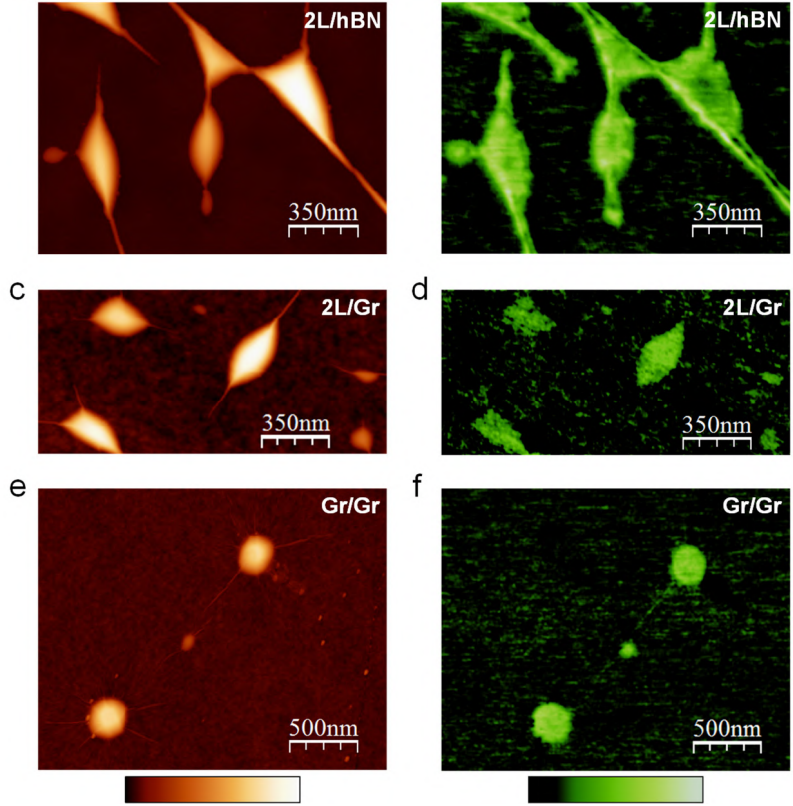

**Figure 3. Control heterostructures: (top) bilayer hBN on thick hBN crystal, (middle) bilayer hBN on graphene; and (lower) graphene on graphene**. (a), (c), (e) Topography of representative regions with bubbles and creases and corresponding (b), (d), (f) electric images (see corresponding dielectric images in Figure S8 in SI). Contrary to the case of monolayer hBN (Figure 1 and 2), no bright areas were detected around or between bubbles in the electric images. Color scales (from dark to bright): topography 20 nm; EFM 1.5 V.

In summary, the experimental evidence that we have presented here clearly indicates the generation of piezoelectric fields in highly strained monolayer hBN. We directly visualized the local electric field generated in the monolayer by strained regions around bubbles and creases, irrespectively of the underlying substrate, while this is absent in bilayer hBN, few-layer hBN and graphene, in agreement with the theory. We calculated the piezoelectric coefficient of hBN, obtaining $\gamma \approx 2.91\times10^{-10}$ Cm$^{-1}$ (0.9 Cm$^{-2}$ when normalized by the layer thickness). This value is comparable to the bulk values of conventional piezoelectric materials such as ZnO, AlN and Lead Zirconate Titanate, (PZT) ceramics.[19] We can estimate the induced polarization and the electric field energy density due to the anisotropic strain gradient in our monolayer hBN membranes. Although they vary strongly, they can reach relatively high levels in some regions, ~$10^{12}$ e$^-$ cm$^{-2}$ and ~$10^{-6}$ eVÅ$^{-2}$, respectively. These are comparable to the carrier concentration in doped graphene[20] and to the energy density in capacitors used in microelectronic circuits[21] if scaled to two-dimensionality. Such strain-induced electric fields can provide a significant scattering mechanism if monolayer hBN is used as encapsulation layer on top of graphene. At the same time, one can envisage that special distribution of the carrier density in graphene can be altered via strain in such monolayer encapsulation layer. Alternative methods to engineer the local strain could be used, such as the use of periodic arrays of nanopillars,[22] and locally control the electric fields. The piezoelectricity of single-layer hBN opens the door to its combination with other 2D crystals for the development of devices with novel functionalities and self-powering potential. These results are also important as they show that electrostatic scanning probe techniques such as electrostatic force microscopy used here or Kelvin probe force microscopy (which simply employs an additional feedback to detect electric variations at the first harmonic, $\omega$) are able to detect piezoelectricity of materials on the nanoscale.

**Experimental Section**

*Samples preparation.* Samples were fabricated using the standard dry transfer technique. Briefly, monolayer hBN was mechanically exfoliated and identified on a double-polymer layer of Polymethylglutarimide (PMGI) and Poly(methyl methacrylate) (PMMA). The PMGI layer was developed from beneath the PMMA layer to create a free-standing and easily-manipulated membrane with the crystal on top. The membrane was then inverted and positioned above the bottom layer (thick hBN or graphene) using a set of micromanipulation stages – with accuracy better than 5 μm. The crystals were then brought into contact. The PMMA was removed by simply peeling back the membrane, meaning no solvent come into contact with either crystal to preserve the cleanliness of the top surface.

*AFM and EFM imaging.* We acquired simultaneous AFM and EFM images using a Nanotec Electronica AFM (see Supporting Information section 1 for details). We measured the EFM force gradient at the first and second harmonic by using a phase-lock loop and a multi-frequency lock-in amplifier (Zurich Instruments). We used n-doped silicon probes (Nanosensors PPP-FMR and PPP-XYNCSTR, mechanical resonance frequencies ~ 65 and 137 kHz and spring constants ~1.8 and 5.3 Nm$^{-1}$ respectively), calibrated using the Sader's method.[23] Doped silicon tips have the advantage of a tip radius of only a few nanometers, one order of magnitude smaller than the typical radius of metal-coated probes, thus increasing the lateral resolution of both topography and EFM images. Furthermore, unlike metal-coated probes, n-doped silicon probes suffer no substantial tip modifications during imaging and, therefore, ensure stable measurement conditions.[24] We oscillated the cantilever in resonance with free amplitudes below 20 nm and setpoints imposing minimum amplitude reduction. We excited the cantilever with an ac voltage of amplitude 4-6 V and frequency 1.8 kHz. These measurement conditions were carefully chosen to minimize all possible sources of cross-talk.[25] We acquired and processed the data using WSxM software.[26] All EFM images in this work are presented with the same scale, contrast and offset for better comparison.

*Theoretical calculations.* For a given strain profile, the induced charge density is obtained from the local variation of the polarization as $\rho(r) = e\, n(r) = -\nabla \cdot P(r) = -\gamma \hat{z} \cdot [\nabla \times \mathcal{A}(r)]$. The numerical calculation involves the following steps (see Supporting Information sections 3 and 4 for details): (i) The equilibrium configuration of a deformed single-layer hBN membrane (61200 or 242000 atoms, depending on the cases, clamped boundary conditions) is obtained from the numerical solution of the discretized elasticity equations for a given shape (*e.g.* circular, triangular or elliptical bubbles). (ii) The solution gives the strain fields $u_{ij}(r)$ generated in the crystal that minimize the energy, which enters in the vector potential $\mathcal{A}(r)$ and which is used to calculate the spatial distribution of electronic polarization $P(r)$, piezoelectric charge density $\rho(r)$, the energy density generated by the piezoelectric effect $u_E(r) = |P(r)|^2/(2\varepsilon_{2d})$, where $\varepsilon_{2d}$ is the dielectric constant of the hBN film.

**Acknowledgements** The authors acknowledge support from EU Graphene Flagship Program (contract CNECTICT-604391), European Research Council Synergy Grant Hetero2D, the Royal Society, Engineering and Physical Sciences Research Council (UK, grant number EP/N010345/1), US Army Research Office (W911NF-16-1-0279). P. A. and L. F. received funding from the EU Research and Innovation Program under the Marie Sklodowska-Curie grant agreement No 793394.

# Supporting Information

**S1. Electrostatic force microscopy (EFM)**

EFM measures the electrostatic force acting between a conductive tip and the sample with an applied electric bias. This force is the sum of the capacitive interaction between the tip and the sample, which depends on its surface potential and dielectric properties of the sample, and the Coulombic interaction between the tip and the static charges/multipoles on the surface, as described in refs. [1]. The total force can then be written as follows:[1a]

$$F = \frac{1}{2}\frac{\partial C}{\partial z}V^2 + E_z Q_t \qquad (1)$$

where $V$ is the electric potential difference and $C$ is the capacitance between the tip and the sample, $Q_t$ is the total charge on the tip, and $E_z$ is the z-component of the electric field arising from the charges on the surface. The capacitive term depends on the first derivative of the tip-sample capacitance, $\partial C/\partial z$, respect to the tip-sample distance, $z$. This in turn is a complicated function of geometrical and dielectric properties of the probe-sample system.[2]

In this work, we carried out ac-EFM imaging by applying an ac bias voltage $V = V_{ac}\sin(\omega t)$ between the tip and the bottom conductive substrate (doped silicon). In this case, it can be shown that the electrostatic force is the sum of three components at 0, $\omega$ and $2\omega$ frequencies

$$F = F_{dc} + F_\omega V_{ac}\sin(\omega t) + F_{2\omega}V_{ac}^2\sin(2\omega t) \qquad (2)$$

where

$$F_\omega = f_\omega\left(C,\frac{\partial C}{\partial z},V_{CPD},\varepsilon\right) + E_{z,dc}C \quad \text{and} \quad F_{2\omega} = f_{2\omega}\left(C,\frac{\partial C}{\partial z},\varepsilon\right) \qquad (3)$$

as described in ref. [1a]. The amplitude of the first harmonic, $F_\omega$, can be written as the sum of a capacitive term and a term proportional to the electric field generated by static charges on the surface. The capacitive term is a complicated function of various parameters, namely, the tip-substrate capacitance and its first derivative, the dielectric properties of the sample, $\varepsilon$, and the tip-substrate contact potential difference, $V_{CPD}$. On the other hand, the amplitude of the second harmonic, $F_{2\omega}$, is not dependent on the static charge distribution on the surface. It is a purely capacitive term that depends only on the tip-substrate capacitance and the dielectric properties of the sample.

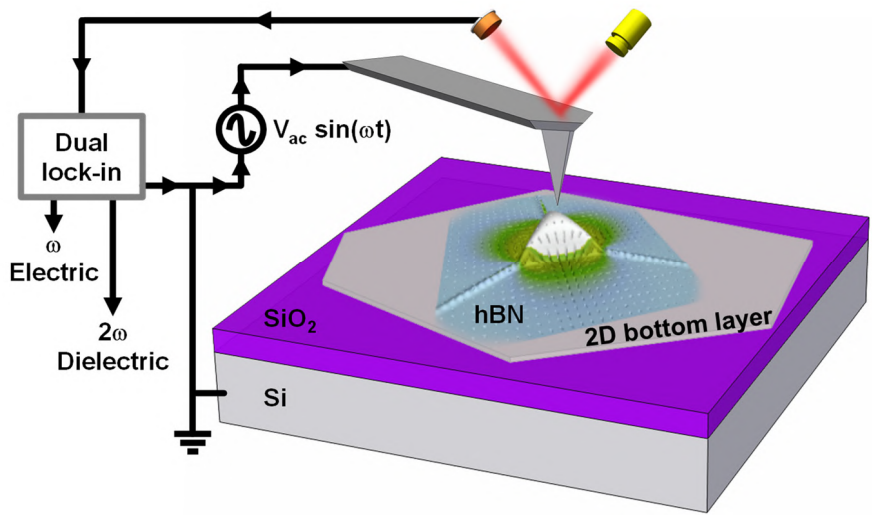

**Figure S1.** Schematic illustration of the experimental setup and the piezoelectric field in monolayer hBN around bubbles. EFM on hBN/2D bottom layer heterostructures on SiO$_2$/Si substrates: An ac voltage bias of frequency $\omega$ is applied between the AFM tip and the Si substrate. Simultaneous electric and dielectric images are measured at the first ($\omega$) and second ($2\omega$) harmonic, respectively. The bubble in the image shows the strain-induced polarization and electric field distribution calculated for a triangular bubble in monolayer hBN.

In this study, we measured both the first harmonic $\omega$ and the second harmonic $2\omega$ using a multifrequency lock-in amplifier (Figure S1). This allowed us to detect the electric field generated by strain-induced charge densities in the $\omega$-image, which we referred to as the electric image. The $2\omega$-image, which we referred to as the dielectric image, is a control image that allowed us to detect any variation in the tip-sample capacitive interaction. We thus verified that the observed variations in the electric images are indeed associated to the presence of static charges and do not reflect variations in the tip-sample capacitive interaction.

Following Glatzel *et al*. Nomenclature,[3] EFM images can be obtained either in amplitude modulation (AM)[4] or in frequency modulation (FM)[5] mode. In AM mode, the electrostatic force is directly detected, while in FM mode it is the force gradient. Here we used the FM mode detecting electric and dielectric images in a two-pass mode, recording the EFM signal at the first and second harmonic simultaneously in the second pass (lift distance 4 nm). The FM mode proves advantageous because it enhances the sensitivity and lateral resolution of the measurement by minimizing the tip-surface distance and stray capacitance contributions. To avoid topographic cross talks, the tip was electrically oscillated with amplitudes of 4-6 V at low frequencies, 1.8 kHz, far from the mechanical vibration of the cantilevers used (at ~ 65 and ~ 137 kHz).

**S2. Additional electrostatic force microscopy images**

**Monolayer hBN on thick hBN crystals**

Figure S2 to S5 below are additional images taken on different regions in monolayer hBN on thick hBN crystals. In particular, the dielectric images corresponding to the electric images of Figure 1a,b in the manuscript are given in Figure S2, revealing no contrast around or between the bubbles, where strain-induced electric contrast is detected. This rules out that such contrast in the electric images is an artefact reflecting a local change in the tip-sample capacitance or in the surface dielectric properties. We found the electric contrast for both triangular and elliptical bubbles and independently of their spatial orientation with respect to the crystallographic axes, in agreement with our theoretical calculations – see sections S3 and S4 below.

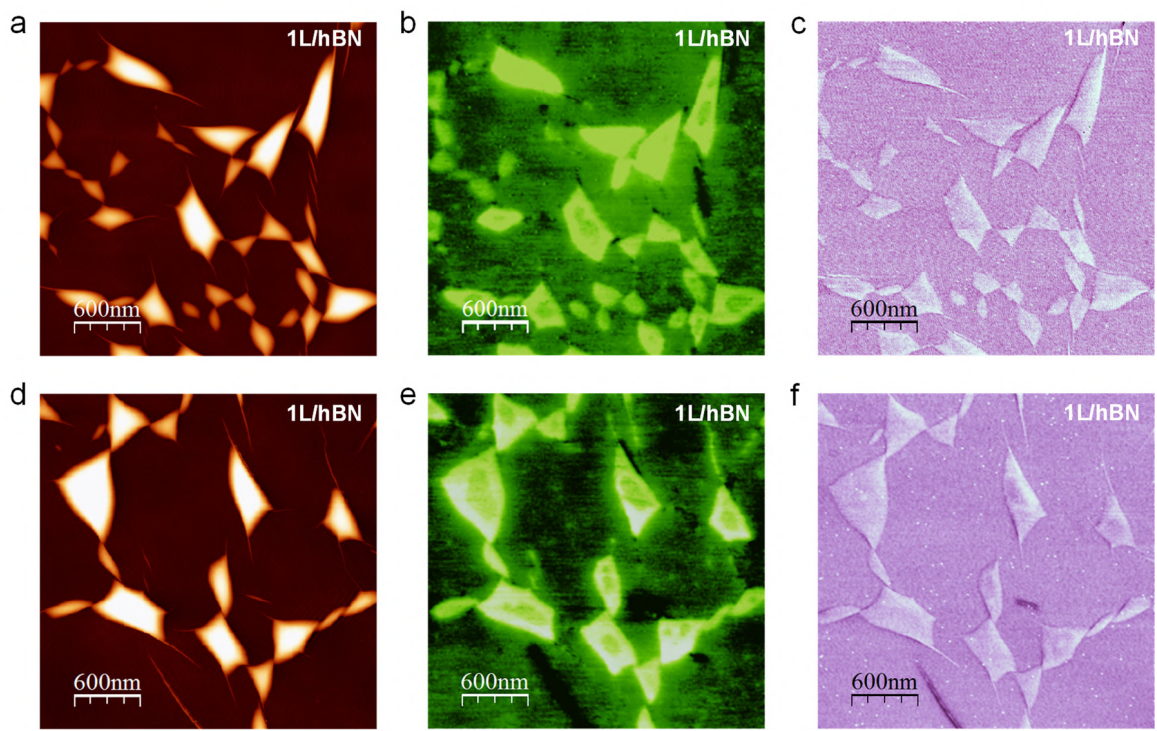

**Figure S2.** Additional AFM and EFM images of regions with bubbles in monolayer hBN on thick hBN crystals on a SiO$_2$/Si substrate. (a), (d) Topography, (b), (e) electric and (c), (f) dielectric images. Bright contrast is detected in flat areas around and between the bubbles in the electric images but not in the dielectric image. Panels (a), (b) correspond to the topography and electric images shown in Figure 1a,b in the main text. . Color scale in panel (b) has been adjusted with respect to Figure 1b to increase the contrast around the smallest bubbles.Color scales (from dark to bright): topography 20 nm; EFM 1 V (b) and 1.5 V (e).

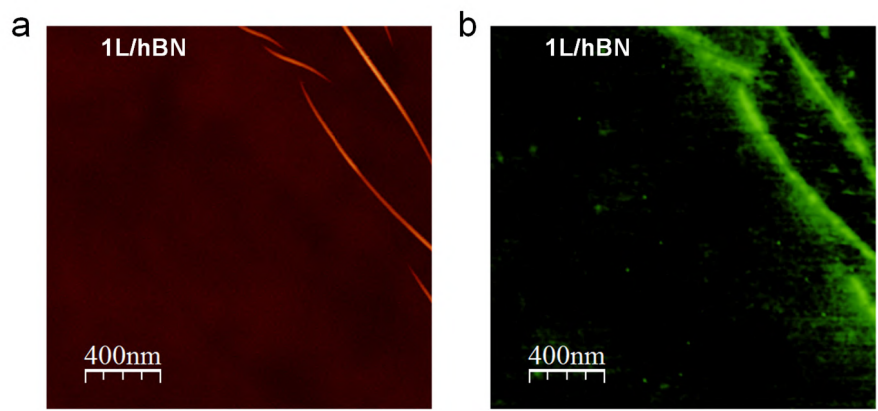

**Figure S3.** AFM and EFM of regions without bubbles in monolayer hBN on thick hBN crystals on a SiO$_2$/Si substrate. (a) Topography and (b) corresponding electric image, showing no bright areas in the flat regions without bubbles. Color scales (from dark to bright): topography 10 nm; EFM 1.5 V.

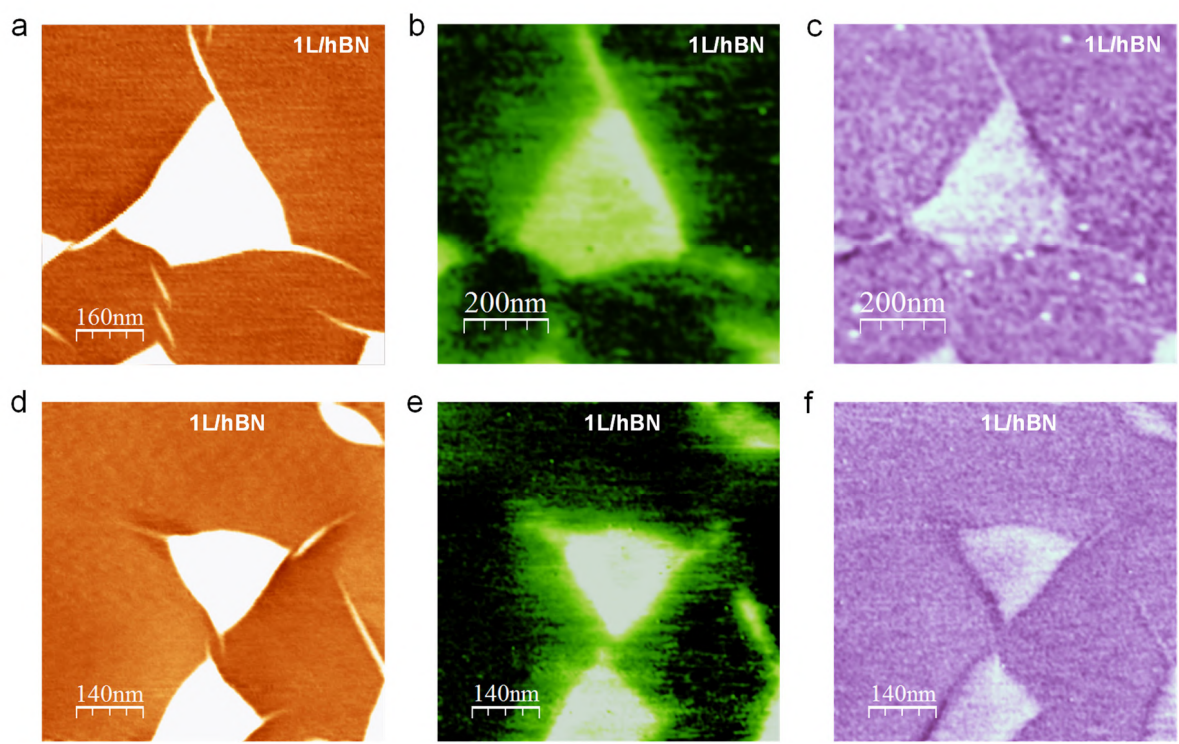

**Figure S4.** AFM and EFM images of triangular bubbles in monolayer hBN on thick hBN crystals on a SiO$_2$/Si substrate. (a), (d) Topography, (b), (e) electric image, showing enhanced contrast around the bubbles and (c), (f) dielectric images, showing no contrast outside bubbles. Panels (a), (b) correspond to the same bubble shown in Figure 1c,d in the main text. Color scales (from dark to bright): topography 2 nm; EFM 1.5 V. Color scale in the topography images was adjusted to show the absence of topographic features as opposed to the electric images.

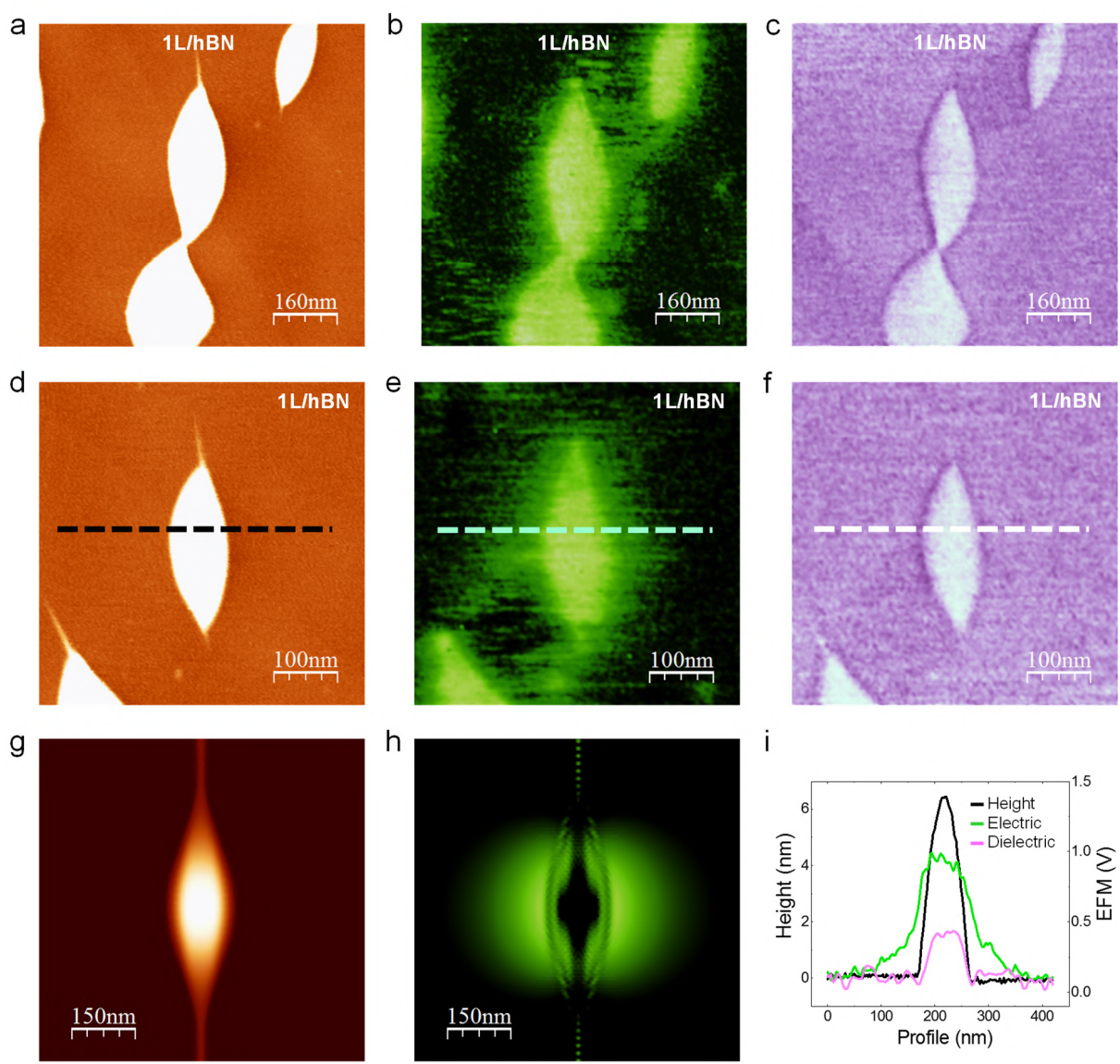

**Figure S5.** AFM and EFM images of elliptical bubbles in monolayer hBN on thick hBN crystals on a SiO$_2$/Si substrate. (a), (d) Topography. (b), (e) Corresponding EFM electric images, showing enhanced contrast around the bubbles. (c), (f) Corresponding EFM dielectric images, showing no contrast around the bubbles. (g) Simulated topography image of an elliptical bubble in monolayer hBN mimicking the experimentally observed bubbles. (h) Corresponding calculated electric-field energy density - see details in sections S3 and S4 (i) Profiles along the lines in (d-f). A clear contrast outside the bubble extending hundreds of nm from the bubble edges is observed only in the electric image (green line). Color scales (from dark to bright): topography 2 nm in (a), (d), 10 nm in (g); EFM 1.5 V; electric-field energy density 1.5 µVÅ$^{-2}$. Color scale in (a), (d) was adjusted to show the absence of features in the topography around the bubbles as opposed to the electric images. The topographic profile in (i) was measured before color-scale adjustment.

**Monolayer hBN on graphene**

Figures S6 and S7 below show additional EFM images taken on monolayer hBN on graphene over Si/SiO$_2$ substrates. Similarly as found for monolayer hBN on thick hBN crystals, bright regions were detected in the electric images in strained regions around and between bubbles and creases, while in such areas no contrast is detected in the dielectric images. In these bright regions, we observed the presence of atomically thin creases (Figure S6g), which generate additional strain and concentrate the electric field between the bubbles. We note that the electric contrast detected in the EFM images around the bubbles in hBN on graphene is generally lower than on thick hBN. We attribute it to the fact that graphene is a semimetal which acts as charge-sink for localized charges. Therefore, enhanced contrast is generally detected in correspondence of ultrathin creases which generate stronger electric fields, as confirmed by our calculations – see the scalebar difference of around one order of magnitude between simulated electric energy densities of bubbles (Figure S16b) and ultrathin creases (Fig S17c).

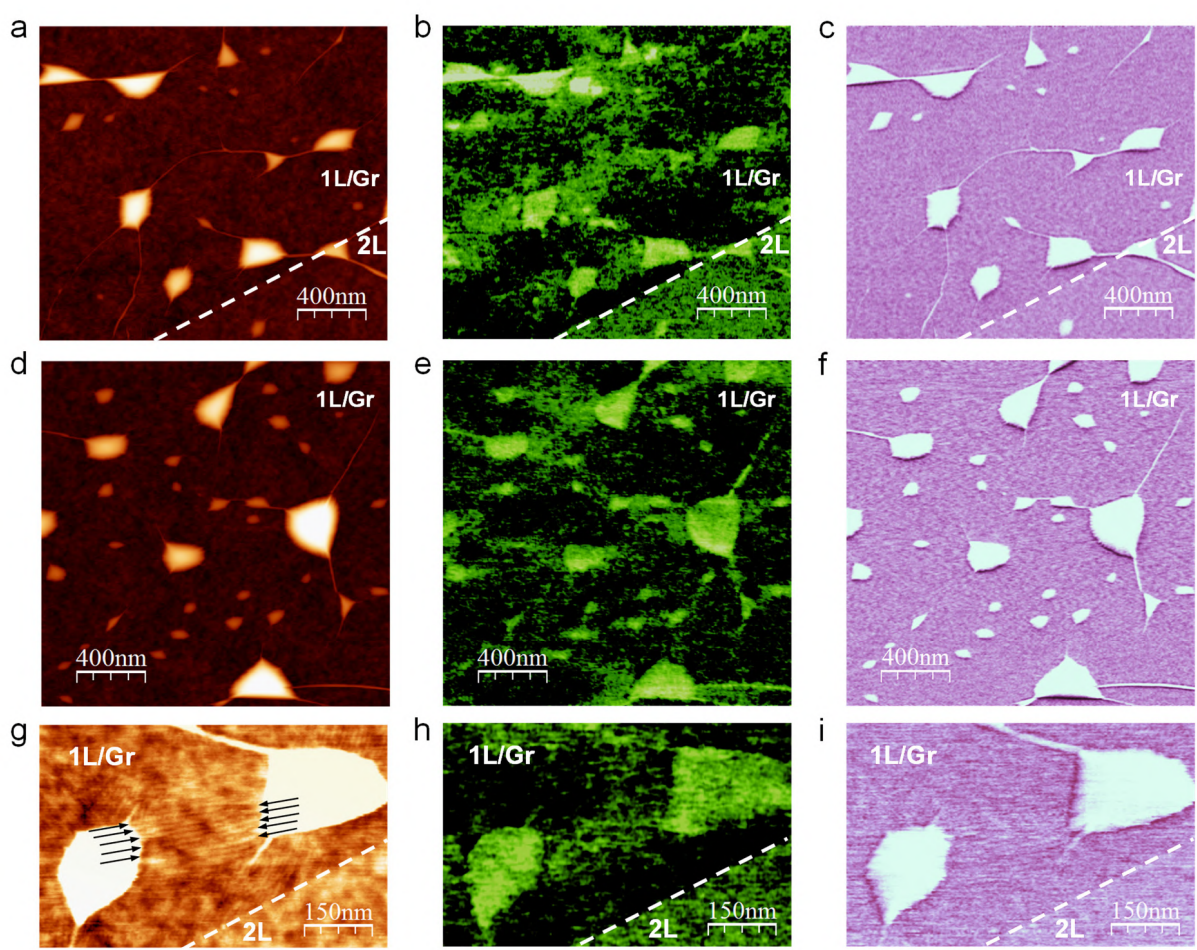

**Figure S6.** Additional AFM and EFM images of regions with bubbles in monolayer hBN on graphene on a SiO$_2$/Si substrate. (a), (d) Topography, (b), (e) electric images, showing enhanced contrast around and between the bubbles, and (c), (f) dielectric images, showing no contrast outside the bubbles. (g) Topography showing the presence of atomically thin creases (marked by black arrows) and (h) corresponding electric image showing enhanced contrast in correspondence of such creases. (i) Dielectric image corresponding to (g), showing no bright contrast in that region. Panels (a), (b) and (g), (h) correspond to the topography and electric images shown in Figure 2a,b and Figure 2c,d in the main text, respectively. Color scales (from dark to bright): topography 20 nm in (a), (d), 2 nm in (g); EFM 1.5 V. Color scale in (g) was adjusted to show the atomically thin creases.

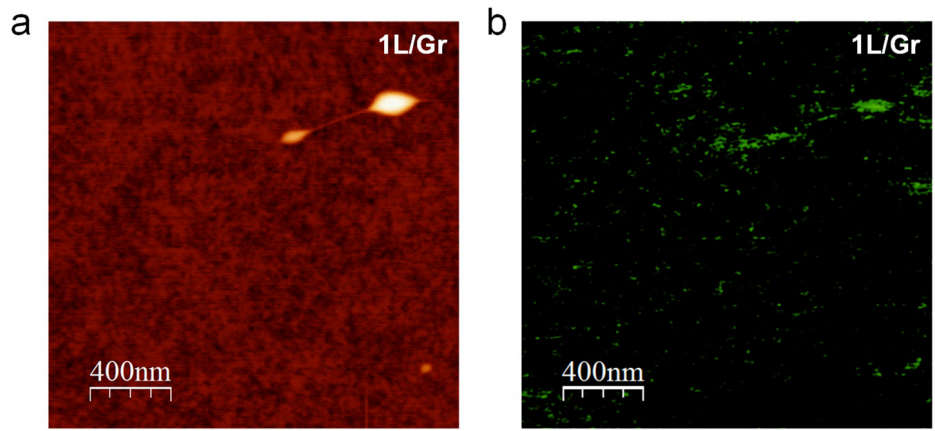

**Figure S7.** AFM and EFM images of a region without bubbles in monolayer hBN on graphene on a SiO$_2$/Si substrate. (a) Topography image and (b) corresponding electric image, showing no areas with enhanced contrast. Color scales (from dark to bright): topography 10 nm; EFM 1.5 V.

**Control samples: bilayer hBN, few-layers hBN and graphene**

Figures S8 to S11 below show EFM images of a series of control samples for which no piezoelectricity is expected, namely, bilayer and few-layers hBN on both thick hBN crystals and graphene, as well as graphene on graphene. Areas with bubbles of similar size and shape as those found in the monolayer hBN were imaged. Contrary to the case of the monolayer hBN, we found no electric contrast in all these layers around or between the bubbles. These images strongly support the piezoelectric origin of the electric contrast that we observed in monolayer hBN. In particular, no contrast was found in correspondance of ultrathin creases in graphene over graphene (Figure S10), which rules out that the strain-induced electric contrast that we observed in monolayer hBN on graphene originates from strain effects in graphene.

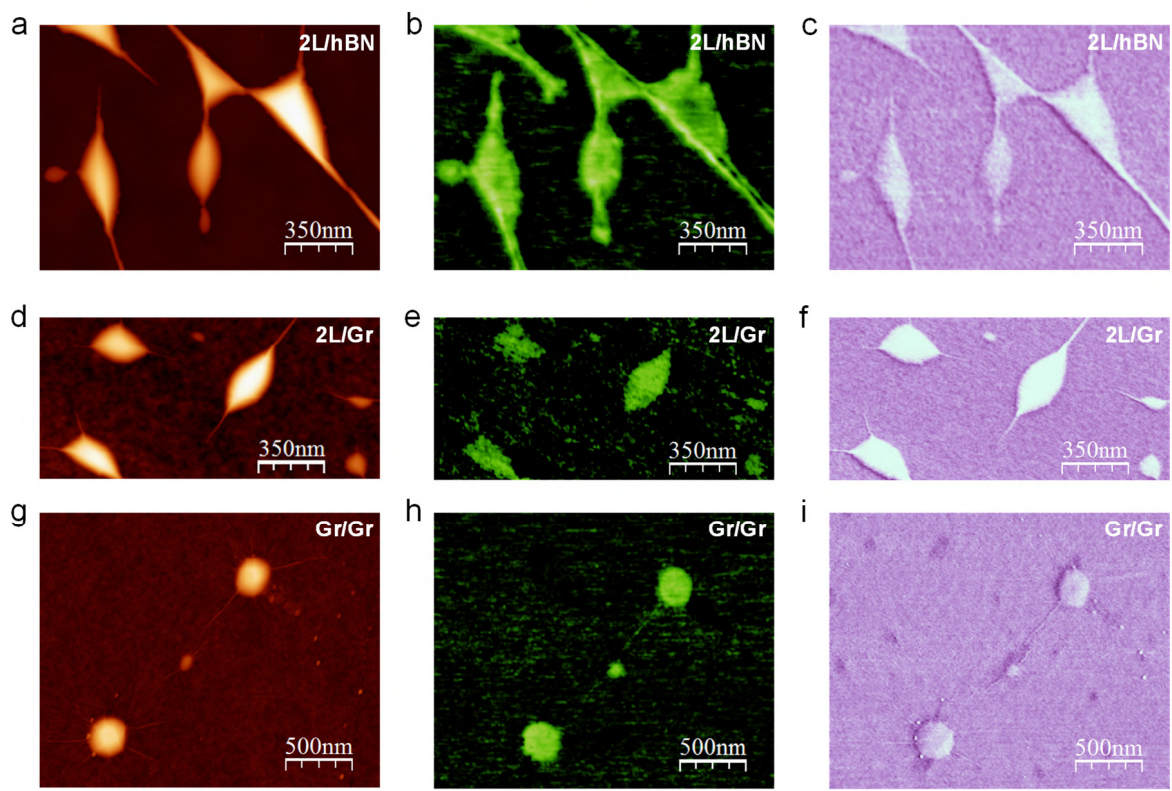

**Figure S8.** AFM and EFM images taken on control heterostructures on a $SiO_2$/Si substrate showing no piezoelectricity, corresponding to Figure 3 in the main text. The dielectric images are also included here. (top) bilayer hBN on thick hBN crystal, (middle) bilayer hBN on graphene; and (lower) graphene on graphene (a), (d), (g) Topography, (b), (e), (h) electric and (c), (f), (i) dielectric images. No bright areas around or between bubbles in the electric images of bilayer hBN (on hBN or graphene) and graphene layers. Color scales (from dark to bright): topography 20 nm; EFM 1.5 V.

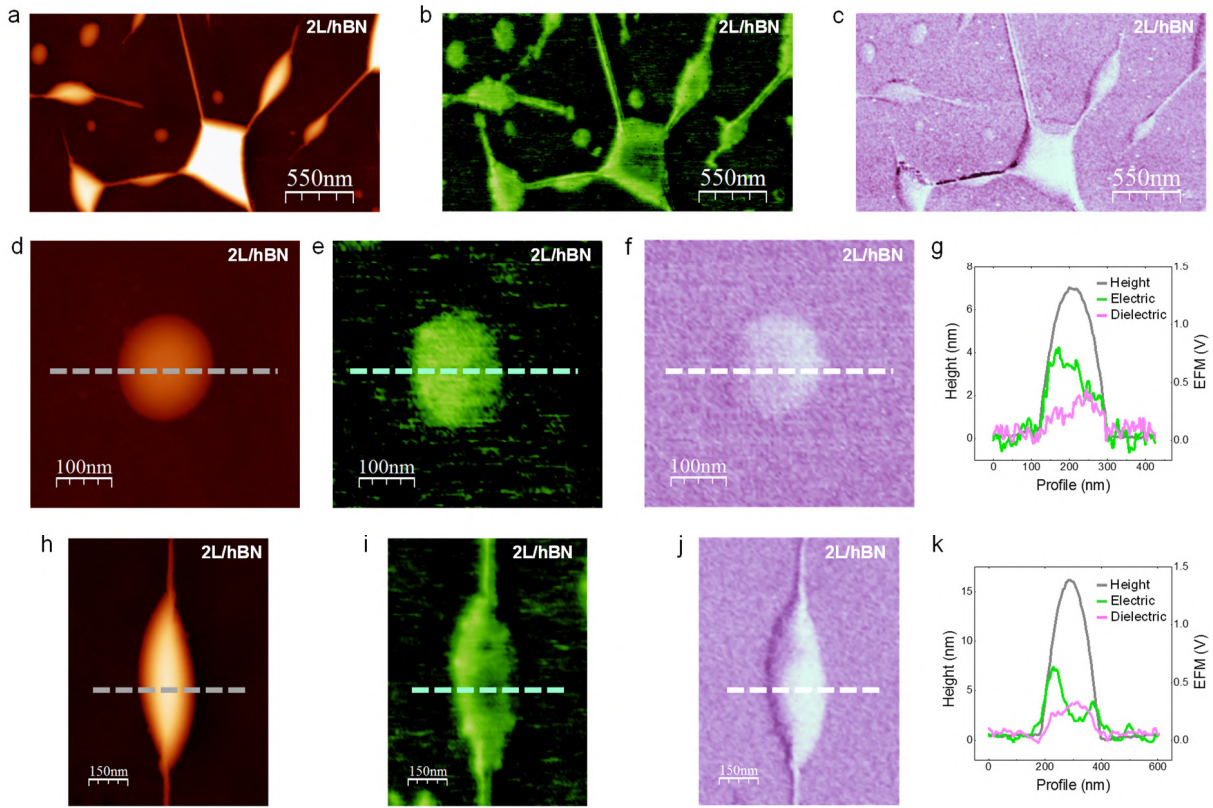

**Figure S9.** AFM and EFM images of different bubbles in bilayer hBN on thick hBN crystals. (a), (d), (h) Topography and corresponding (b), (e), (i) electric and (c), (f), (j) dielectric images, showing no bright areas around or between the bubbles. (g), (k) Profiles along the lines in (d-f) and (h-j) respectively, where the absence of electric contrast outside the bubble s (green line) is clearly visible as opposed to the case of the monolayer hBN (see profile in Figure 1e in the main text). Color scales (from dark to bright): topography 20 nm; EFM 1.5 V.

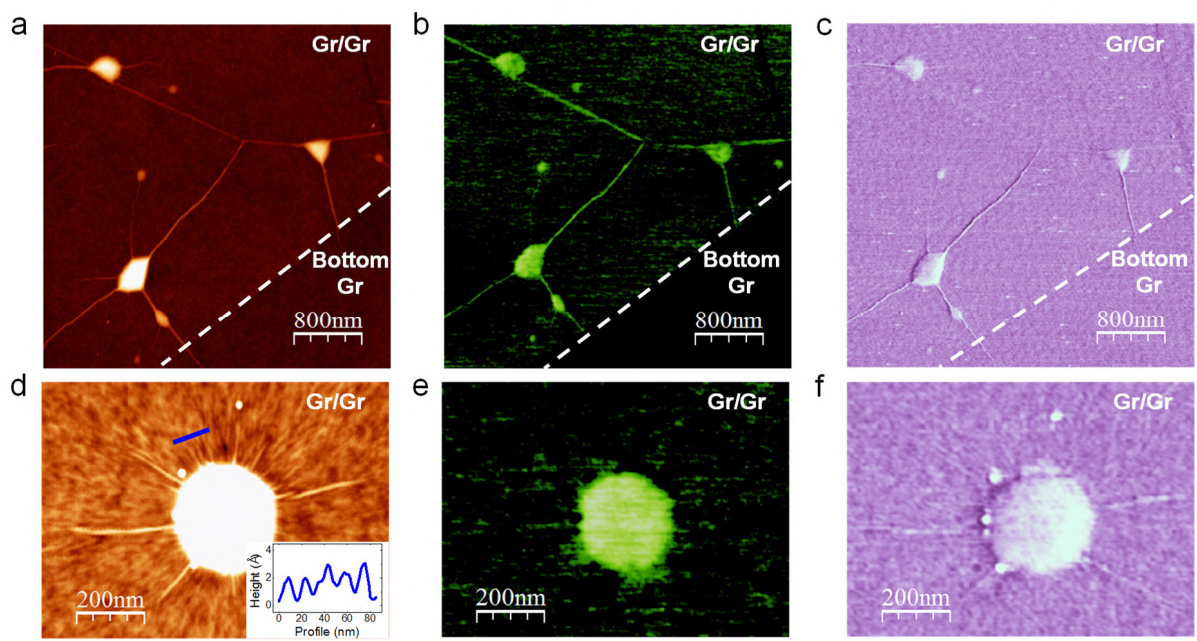

**Figure S10.** AFM and EFM images of bubbles in graphene on graphene. (a) Topography and corresponding (b) electric and (c) dielectric images, showing no bright areas around or between the bubbles. (d) Topography showing atomically thin creases around a bubble in graphene. The inset is a topographic profile along the blue line. Corresponding electric (e) and dielectric (f) images showing no electric contrast in correspondence of such ultrathin creases. Color scales (from dark to bright): topography 20 nm in (a), 2 nm in (d); EFM 1.5 V. Color scale in (d) was adjusted to show the atomically thin creases.

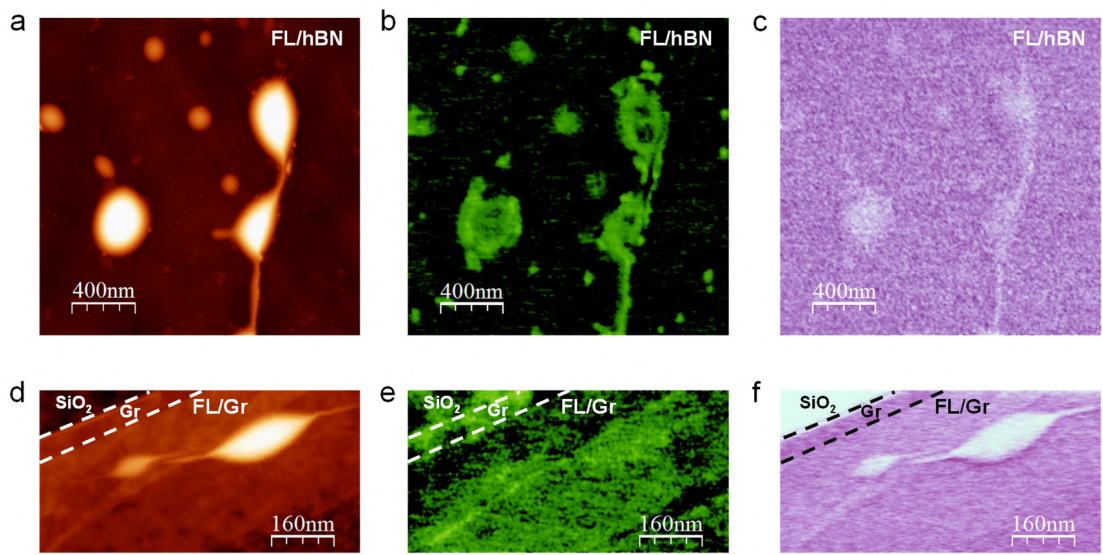

**Figure S11.** AFM and EFM images of bubbles in few layer hBN on hBN (top row) and on graphene (bottom row). (a), (d) Topography and corresponding (b), (e) electric and (c), (f) dielectric images, showing no bright areas around or between the bubbles. Color scales (from dark to bright): topography 10 nm; EFM 1.5 V.

**S3. The elasticity problem in the deformed honeycomb lattice**

Here we describe the method used to compute the piezoelectric induced polarization and charge density generated by bubbles in monolayer hBN. Our technical analysis consists in solving the elasticity equations for the membrane in the honeycomb lattice by means of a discretization procedure. The strain fields obtained for different profiles of bubbles are used to calculate the piezoelectric induced charge density and the electric fields.

We start by considering a single sheet of hBN in the continuum limit, in which the free energy of the membrane is given by the classical theory of elasticity:[6]

$$F_{el}[\mathbf{u}, w] = \frac{1}{2} \int d^2 x [\kappa (\nabla^2 w)^2 + (\lambda u_{ii}^2 + 2\mu u_{ij}^2)] \quad (4)$$

where $u_{ij}$ is the strain tensor

$$u_{ij} = \frac{1}{2}(\partial_i u_j + \partial_j u_i + \partial_i w \partial_j w) \quad (5)$$

with $u_i$ the field associated to the in plane displacements in the $i-th$ direction ($i = x, y$), $w$ that associated to the out of plane displacements, $\kappa = 0.82$ eV is the bending rigidity, $\lambda = 23.39$ eV/$l^2$ and $\mu = 49.55$ eV/$l^2$ are the Lamé coefficients of hBN,[7] $l$ being the lattice constant. The value of $l$ in the atomic limit is $l_0 = \sqrt{3} a_0 = 2.52$ Å, which gives: $\lambda_0 = 3.68$ eVÅ$^{-2}$ and $\mu_0 = 7.80$ eVÅ$^{-2}$.

We consider the case in which the configuration of the out of plane displacements is defined as to reproduce the different shapes of bubbles observed in the experiments, $w(x, y) \equiv w_0(x, y)$. Therefore the free energy $F_{el}$ is a functional of the in plane fields $u_i$ only, and the bending energy represented by the first term of equation (4) can be safely neglected, since it is just an additive constant to the total energy.

At equilibrium, the configuration of the $u_i$ fields that minimizes $F_{el}$ is given by the solution of the following elasticity equations:

$$\lambda \partial_x \left( \partial_x u_x + \partial_y u_y + \frac{|\nabla w|^2}{2} \right) + \mu \partial_x [2 \partial_x u_x + (\partial_x w)^2] + \mu \partial_y (\partial_y u_x + \partial_x u_y + \partial_x w \partial_y w) = 0 \quad (6a)$$

$$\lambda \partial_y \left( \partial_x u_x + \partial_y u_y + \frac{|\nabla w|^2}{2} \right) + \mu \partial_y [2 \partial_y u_y + (\partial_y w)^2] + \mu \partial_x (\partial_y u_x + \partial_x u_y + \partial_x w \partial_y w) = 0 \quad (6b)$$

that are nothing but the static Euler-Lagrange equations.

We next generalize equations (6) to the discrete case by replacing the partial derivatives with their corresponding finite differences on the honeycomb lattice. For this we follow the approach of refs [8]. The honeycomb lattice consists of two sub-lattices, that here we label *A* and *B* (Figure S12). In the particular case of hBN, the two species correspond to Nitrogen and Boron atoms, respectively. Each atom of type *A* has three first nearest neighbors of type *B*, that we labeled with the indices 1,2,3, and six second nearest neighbors of type *A*, labeled by the indices 4,…9.

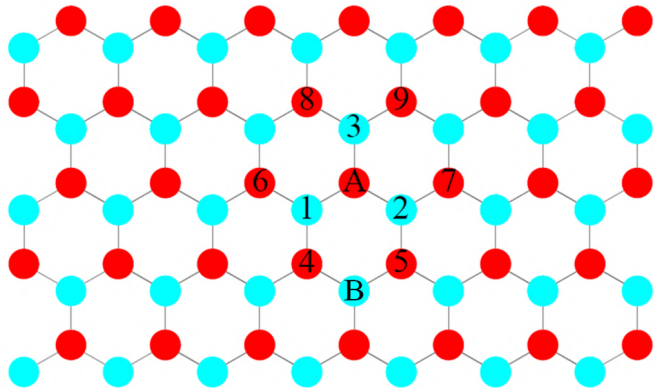

**Figure S12.** The honeycomb lattice of hBN. For a given site of type *A* (red), there are three first nearest neighbors 1,2,3 of type *B* (cyan) and six second nearest neighbors 4,…9 of type *A*.

According to Figure S12, if the point $A$ has coordinates $(x, y)$, the coordinates of its nine nearest neighbors are then:

$$n_1 = \left(x - \frac{l}{2}, y - \frac{l}{2\sqrt{3}}\right), \quad n_2 = \left(x + \frac{l}{2}, y - \frac{l}{2\sqrt{3}}\right), \quad n_3 = \left(x, y + \frac{l}{\sqrt{3}}\right)$$

$$n_4 = \left(x - \frac{l}{2}, y - \frac{l\sqrt{3}}{2}\right), \quad n_5 = \left(x + \frac{l}{2}, y - \frac{l\sqrt{3}}{2}\right), \quad n_6 = (x - l, y)$$

$$n_7 = (x + l, y), \quad n_8 = \left(x - \frac{l}{2}, y + \frac{l\sqrt{3}}{2}\right), \quad n_9 = \left(x + \frac{l}{2}, y + \frac{l\sqrt{3}}{2}\right) \quad (7)$$

An analogous scheme holds for each atom of type $B$, with an equivalent definition of the nearest neighbors. The partial derivatives of first and second order that appear in the continuum equations (6) can be replaced by their corresponding finite differences on the lattice, by introducing the following operators:[8]

$$Tf(A) = f(n_1) - f(A) + f(n_2) - f(A) + f(n_3) - f(A) \sim \frac{l^2}{4}\nabla^2 f \quad (8a)$$

$$Hf(A) = f(n_6) - f(A) + f(n_7) - f(A) \sim l^2 \partial_x^2 f \quad (8b)$$

$$Df(A) = f(n_4) - f(n_5) + f(n_9) - f(n_8) \sim l^2 \sqrt{3} \partial_x \partial_y f \quad (8c)$$

$$\Delta_x f(A) = \frac{f(n_7) - f(n_6)}{2} \sim l \partial_x f \quad (8d)$$

$$\Delta_y f(A) = \frac{f(n_3) - f(A) - [f(n_1) - f(A) + f(n_2) - f(A)]}{2} \sim \frac{l}{\sqrt{3}} \partial_y f \quad (8e)$$

where $f$ is a generic function of the lattice positions. Thus, the elasticity equations on the honeycomb lattice can be written as:

$$4\mu T u_x + (\lambda + \mu) H u_x + \frac{\lambda + \mu}{\sqrt{3}} D u_y + \frac{\lambda + \mu}{l}\left[\Delta_x w H w + \Delta_y w D w\right] + \frac{4\mu}{l} \Delta_x w T w = 0 \quad (9a)$$

$$4(\lambda + 2\mu) T u_y - (\lambda + \mu) H u_y + \frac{\lambda + \mu}{\sqrt{3}} D u_x + \frac{4\sqrt{3}}{l}(\lambda + 2\mu)\Delta_y w T w + \frac{\lambda + \mu}{l\sqrt{3}}\left[\Delta_x w D w - 3\Delta_y w H w\right] = 0 \quad (9b)$$

Notice that in equations (8d) and (8e) we used a symmetric definition of the first order differences, which differs from the one adopted in previous works.[8] Although this choice is completely irrelevant when dealing with periodic boundary conditions, it becomes necessary in the case of clamped boundary conditions (zero displacements at the edges of the lattice), in order to prevent the appearance of any preferential direction.

As an illustrative example, Figure S13 shows the equilibrium configuration of the membrane as obtained from the numerical solution of equations (9) in the presence of clamped boundary conditions and of a circular bubble given by: $w(x, y) = w_{max}\left[1 - \left(\frac{r}{R}\right)^2\right]\Theta\left[1 - \left(\frac{r}{R}\right)^2\right]$, with $r^2 = x^2 + y^2$, $R = 4l$, and $w_{max} = 4.33l$. As it is evident from the stretching (and eventually shrinking) of the B-N bonds in proximity of the bubble, the system is forced to develop a finite strain $u_{ij}$, that would be absent in the flat case $w = 0$.

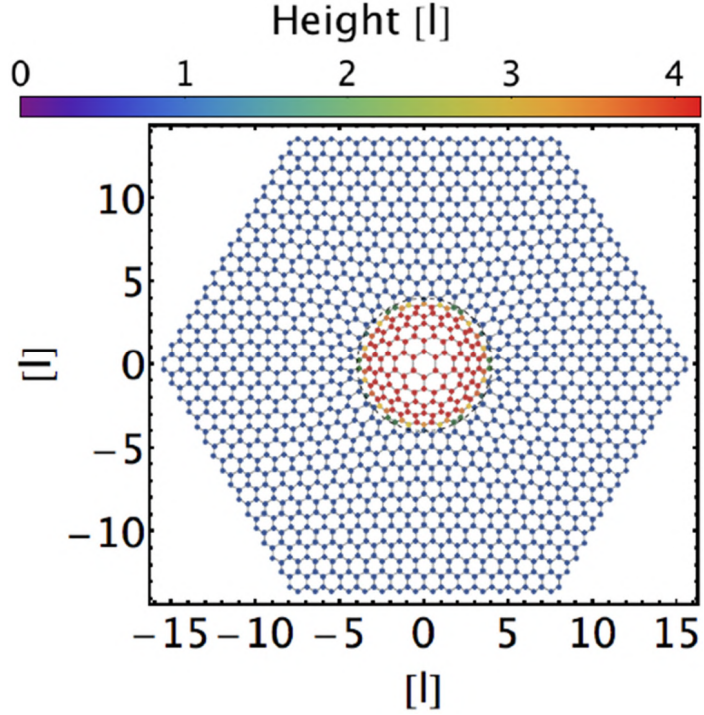

**Figure S13.** Example of circular bubble. Numerical solution of the elasticity equations (9) in the presence of clamped boundary conditions and of a circular bubble. The out of plane displacement $w$ is represented as a color map displaying the heights of the lattice points in units of $l$.

## S4. Evaluation of the charge density and electric field generated by the piezoelectric effect in the presence of bubbles

In the presence of a finite strain field $u_{ij}$, piezoelectric crystals acquire a polarization $P$ given by: $P_i = \sum_{jk} \gamma_{ijk} u_{jk}$, where $\gamma_{ijk}$ is the third-rank piezoelectric tensor. The local variations of this polarization field can generate charge *self-doping*, with charge density given by:[9]

$$\rho = -\nabla \cdot \boldsymbol{P} \tag{10}$$

Equation (10) implies that $\boldsymbol{P}$ is proportional to the electric field $\boldsymbol{E}$, according to the Maxwell law: $\boldsymbol{E} = -\boldsymbol{P}/\varepsilon_{2d}$, where $\varepsilon_{2d} \equiv \varepsilon_r \varepsilon_0 l_\perp$ is the dielectric constant of the bi-dimensional film, with $\varepsilon_r$ the relative permittivity along the $c$ axis, $\varepsilon_0$ the permittivity of the vacuum, and $l_\perp$ the interlayer spacing. The characterizations of hBN given by the refs. [10] provide the following values: $\varepsilon_r = 6.85$ and $l_\perp = 3.33$ Å.

For the case of monolayer hBN and related 2D crystals, one can exploit the symmetries of the piezoelectric tensor to write the piezoelectric induced polarization as:[11]

$$\boldsymbol{P} = \gamma \mathcal{A} \times \hat{\boldsymbol{z}} \tag{11}$$

with

$$\mathcal{A} = (u_{xx} - u_{yy})\hat{\boldsymbol{x}} - 2u_{xy}\hat{\boldsymbol{y}} \tag{12}$$

and $\gamma$ the piezoelectric constant of the material. Here we use the value $\gamma = 2.91 \times 10^{-10}$ Cm$^{-1}$, as given by the analytic estimate based on the $\boldsymbol{k} \cdot \boldsymbol{p}$ method.[11]

Using equations (10)-(12) along with the definition (5) of the strain field $u_{ij}$, the charge density can be written as:

$$\rho = \gamma[(\partial_x^2 - \partial_y^2)u_y + (\partial_y w)(\partial_x^2 - \partial_y^2)w + 2(\partial_x \partial_y u_x + \partial_x w \partial_x \partial_y w)] \tag{13}$$

In the case of the honeycomb lattice that we are considering here, the partial derivatives appearing in the previous expression must be replaced by their corresponding finite differences defined in the equations (8), which gives:

$$\rho = \frac{\gamma}{l^2}\left[2(H-2T)u_y + \frac{2\sqrt{3}}{l}\Delta_y w(H-2T)w + \frac{2}{\sqrt{3}}\left(Du_x + \frac{1}{l}\Delta_x w Dw\right)\right] \quad (14)$$

**S5. Numerical results**

Here we report on the numerical results obtained by solving the discrete elasticity equations (9) in the presence of different bubble shapes. In particular, given the polarization field $\boldsymbol{P}$, we compute the charge density $\rho$ according to equation (14) and the energy density of the electric field generated by the piezoelectric effect: $u_E = \frac{\varepsilon_{2d}}{2}|\boldsymbol{E}|^2 = \frac{1}{2\varepsilon_{2d}}|\boldsymbol{P}|^2$. The results refer to lattices of size $N_s = 61200$ atoms (or bigger, when specified) with clamped boundary conditions. From the energy density of the electric field generated by the piezoelectric effect we can obtain the electrostatic force distribution acting normal to the surface by computing the first derivative with respect to the out of plane direction.[12]

Figure S14 shows the numerical results obtained in the case of a perfectly circular bubble, as given by the following height profile:

$$w(x,y) = w_0 e^{-\left(\frac{x^2+y^2}{R^2}\right)} \quad (15)$$

which displays a gaussian decay moving from the center of the bubble. Here $w_0 = 28$ nm is the maximum height of the bubble and $R = 380$ nm its radius.

Figure S15 displays a triangular bubble having the same shape of that of Figure 1f in the main text, but different orientations $\theta$ with respect to the crystallographic axes. We considered the cases $\theta = 0°$ (first column), $\theta = 20°$ (second column), $\theta = 40°$ (third column), and $\theta = 60°$ (fourth column). These cases show that the distribution of energy and electric fields is not strikingly affected by the particular orientation of the bubble, but rather by its geometry, as the energy results to be concentrated mainly along the sides of the bubble and the creases.

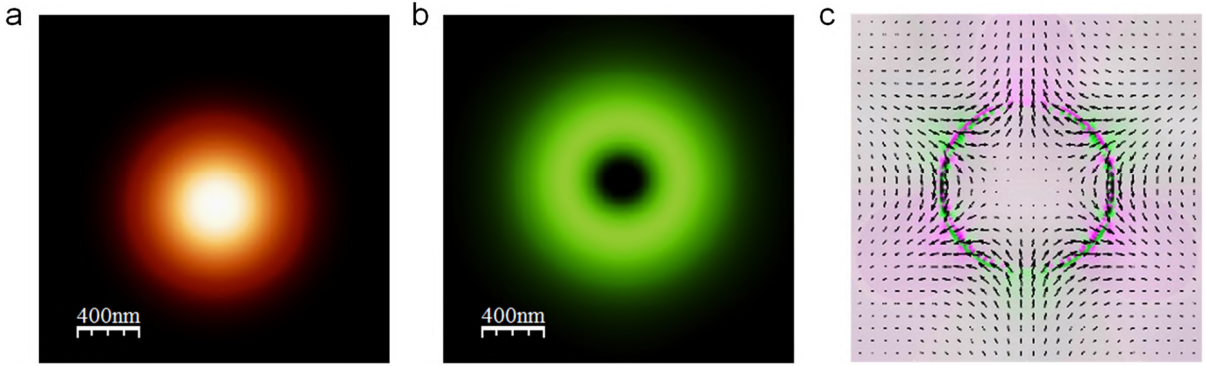

**Figure S14.** Perfectly circular bubble with height profile given by equation (15). (a) Topography. (b) Energy density of the electric field. (c) Charge density. The superimposed arrows represent the polarization field. Color scales (from dark to bright): topography 30 nm; electric field energy density 2 μVÅ$^{-2}$; charge distribution -3×10$^{12}$ (green) to 3×10$^{12}$ (pink) e$^-$ cm$^{-2}$.

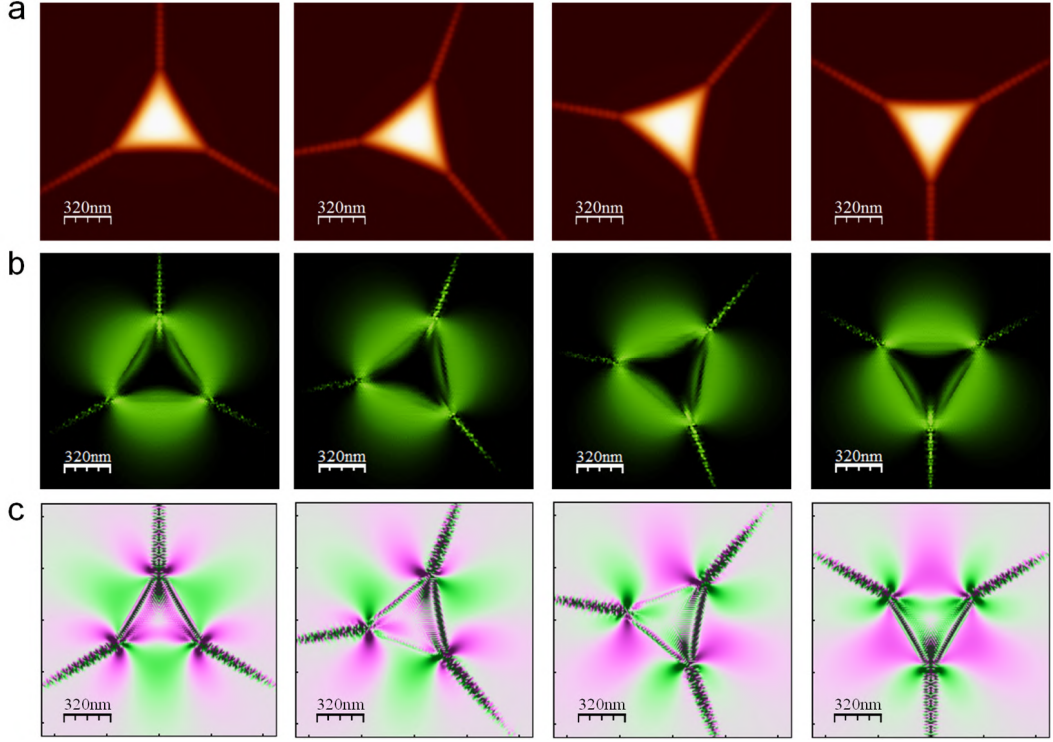

**Figure S15.** Triangular bubble at different orientations with respect to the crystallographic axes. (a) Topography. (b) Energy density. (c) Charge distribution. Columns from left to right correspond to rotations of $\theta = 0°, 20°, 40°$ and $60°$ with respect to the crystallographic axes, respectively. Color scales (from dark to bright): topography 20 nm; electric field energy density 1.5 µVÅ$^{-2}$; charge distribution -5×10$^{12}$ (green) to 5×10$^{12}$ (pink) e$^-$ cm$^{-2}$.

Figure S16 refers to the geometrical configuration of an elliptical bubble with one-dimensional creases at the vertical corners. This bubble shape has been obtained by means of the following analytical expression for the height field:

$$w(x,y) = w_0 exp\left[-\left(\frac{x}{b}e^{y^2/c^2}\right)^4\right]e^{-2y^2/c^2} + w_c e^{-x^2/\xi^2}\left[e^{-\left(\frac{y-2c}{c}\right)^2} + e^{-\left(\frac{y+2c}{c}\right)^2}\right] \quad (16)$$

where $w_0 = 10$ nm is the maximum height of the bubble, $b = 57$ nm, $c = 210$ nm, $w_c = 1$ nm is the maximum height of each crease and $\xi = 10$ nm sets the width of the creases.

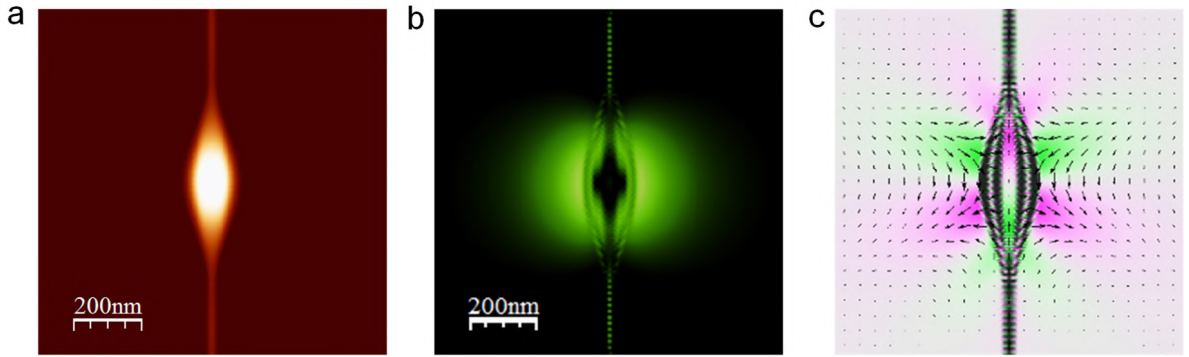

**Figure S16.** Elliptical bubble with one-dimensional creases as obtained from the height profile given by equation (16). (a) Topography. (b) Energy density of the electric field. (c) Charge density. The superimposed arrows represent the polarization field. (d) Electrostatic force distribution. Color scales (from dark to bright): topography 10 nm; electric field energy density 2 µVÅ$^{-2}$; charge distribution -1×10$^{12}$ (green) to 1×10$^{12}$ (pink) e$^-$ cm$^{-2}$.

As described in the main text, in the case of monolayer hBN on graphene we found atomically thin creases around many bubbles, which generate large strains. Such creases can extend over distances up to hundreds of nanometers from the bubbles and connect different bubbles. To better compare our experimental images with the simulations, we included such ultrathin creases in the theoretical model. We used the following analytical expression to modulate the height field in the configuration in which two bubbles are connected by many ultrathin creases, as in the experimental observation of Figure 2c-h in the main text:

$$w(x,y) = w_0 e^{-2y^2/c^2}\left\{exp\left[-\left(\frac{x-D/2}{b}e^{y^2/c^2}\right)^4\right] + exp\left[-\left(\frac{x+D/2}{b}e^{y^2/c^2}\right)^4\right]\right\} +$$
$$+ \sum_{n=-N}^{N} w_c e^{-[(y-nd)/\xi]^2}\left\{e^{-(1-2x/D)^2} + \theta(x+D/2)e^{-(1+2x/D)^2} + [1-\theta(x+D/2)]e^{-[(x+D/2)2/b]^2}\right\} \quad (17)$$

where $w_0 = 20$ nm is the maximum height of each bubble, $D = 400$ nm is the distance between their centers, $b = 75$ nm, $c = 150$ nm, $2N + 1 = 11$ is the number of creases, $d = 10$ nm is the distance between the creases, $w_c = 0.45$ nm sets the height of the creases in $x = 0$ at approximatively 0.3-0.4 nm, and $\xi = 2.5$ nm sets the width of each crease. The numerical results concerning this configuration have been obtained by solving the elasticity equations in a lattice of $N_s = 242400$ atoms. Such a large lattice size was necessary to reproduce the narrow features of the creases, as shown in Figure S17 below. Figure S17a shows the simulated topography including atomically thin creases that connect two elliptical bubbles. As it can be seen in the topographic profile (Figure S17b), the simulation reproduces the spacing (~ 10 nm) and the height (~ 3 Å) of the creases that we experimentally observed – see the experimental topography in Figure 2c in the main text. Figure S17c,d shows the corresponding calculated electric-field energy density and its profile. It predicts that the piezoelectric field extends along the ultrathin creases in the region between the bubbles, in good agreement with our observations in the EFM electric image (Figure S17e,f and Figure 2d in the main text).

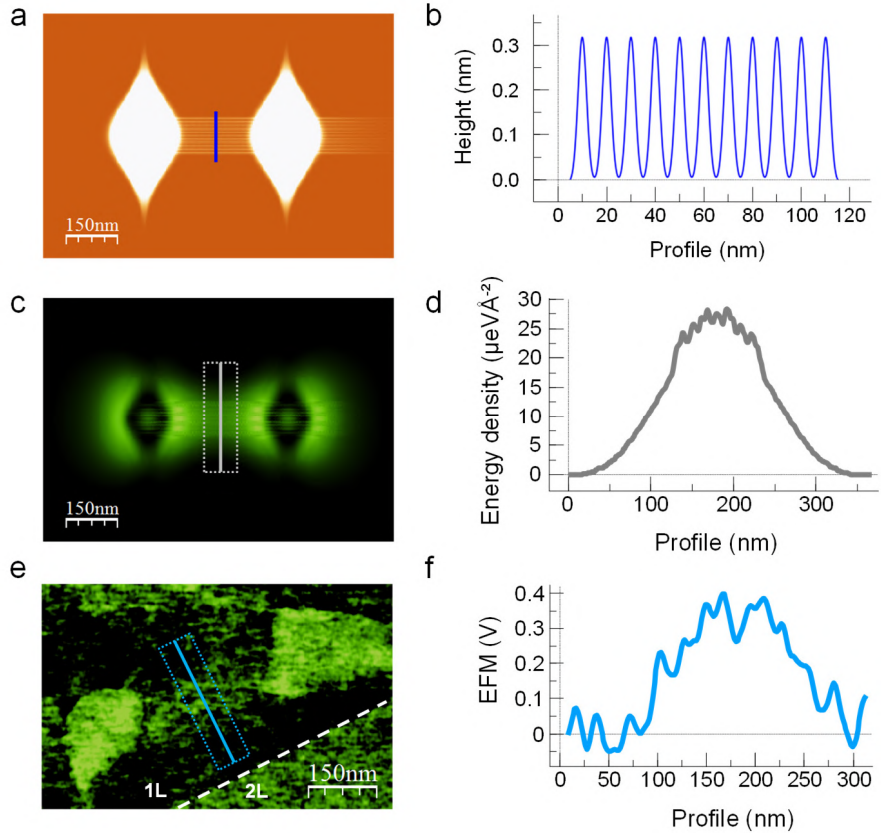

**Figure S17.** Analytical simulation of thin creases mimicking the experimental observations. (a) Simulated topography image, same image as Figure 2f in the main text, with height profile given by equation (17). (b) Profile along the line in (a). (c) Electric field energy density simulation, same image as Figure 2g in the main text. (d) Average smoothed profile along the vertical line and in (c) corresponding to the area enclosed by the dotted rectangle. (e) Experimental EFM electric data, same image as Figure 2d in the main text. (f) Average smoothed profile along the line in (e) corresponding to the area enclosed by the dotted rectangle.